\documentstyle[preprint,aps,prb]{revtex}       

\begin{document}
\draft
\date{\today}
\title{ Localization and Fluctuations in Quantum Kicked Rotors }
\author{Indubala I. Satija\footnote{email:isatija@sitar.gmu.edu}}
\address{
 Department of Physics, George Mason University,
 Fairfax, VA 22030}
\author{ Bala Sundaram\footnote{email:bas@math.csi.cuny.edu} }
\address{ Department of Mathematics, 1S215
CSI-CUNY
2800, Victory Boulevard
Staten Island, NY 10314}
\author{ Jukka A. Ketoja}
\address{KCL Paper Science Centre, P.O. Box
70, FIN-02151, Espoo, Finland}
\date{\today}
\maketitle
\tightenlines
\begin{abstract}

We address the issue of fluctuations, about an exponential
lineshape, in
a pair of one-dimensional kicked quantum systems exhibiting
dynamical localization. 
An exact renormalization scheme establishes the fractal
character of the fluctuations and provides a new method to
compute the localization length in terms of the fluctuations.
In the case of a linear rotor, the fluctuations
are independent of the kicking parameter $k$ and
exhibit self-similarity for certain values of the quasienergy. For
given $k$, the asymptotic localization
length is a good characteristic of the localized lineshapes for all
quasienergies. This is in stark contrast to the
quadratic rotor, where
the fluctuations depend upon the strength of the kicking and 
exhibit local "resonances". These resonances result in
strong deviations of the localization length from
the asymptotic value. The consequences are particularly pronounced when
considering the time evolution of a packet made up of several
quasienergy states. 

\end{abstract}

\narrowtext

Dynamical localization is an important manifestation of the quantum
suppression of diffusive classical motion resulting from nonintegrable 
dynamics~\cite{Reichl,Casati,Fishman,Fishman2,Izrailev}. As the name suggests, the mechanism is analogous to the
Anderson description of a low-dimensional, low-temperature insulator
phase in terms of
tight-binding models (TBM)~\cite{Anderson,TVR}. The relationship between these two
seemingly disparate systems was made explicit~\cite{Fishman} in a class of kicked
quantum Hamiltonians of the form (in dimensionless units):
\begin{equation}
H=K(p) + V(\theta) \sum \delta(t-n) \; ,
\end{equation}
where $K(p)$ denotes a general kinetic energy operator. Note that time 
is measured in units of spacing between kicks. The
time-periodic nature of the Hamiltonian allows the time-dependent
solution to be expressed,
in terms of the one-step evolution operator $U$, as
\begin{equation}
U\psi(\theta,t)= \psi(\theta, t+1)\;.
\end{equation}
For kicked systems, $U$ takes on the particularly simple form:
\begin{equation}
U= \exp{(-iK(p)/\hbar)} \exp{(-iV(\theta)/\hbar)}\;,
\end{equation}
where $\hbar$ refers to the effective quantization scale in
dimensionless units. 
Further, the evolution of any initial condition can be expressed in
terms of quasienergy states $\phi_\omega$ which satisfy
\begin{equation}
U\phi_\omega = e^{-i\omega} \phi_\omega \;,
\end{equation}
where the quasienergy $\omega$ is real as $U$ is an unitary operator.

The relationship between these quantum kicked systems and TBM
becomes clear on projecting the quasienergy states onto eigenstates of 
$K(p)$. In our context, these are angular momentum states. The
equation satisfied by the projection coefficients $u_m$ onto the
$m^{th}$ angular momentum state is~\cite{Fishman}
\begin{equation}
T_m u_m + \sum_r W_{n-r} u_r = 0 \;,
\end{equation}
where
\begin{equation}
T_m = \tan{\left[ (\omega - K(m))/2 \right]},
\end{equation}
and the $W_m$ are the Fourier weights of 
\begin{equation}
W(\theta)=-\tan{(V(\theta)/2\hbar)} \;,
\end{equation}
with respect to the angular momentum basis. This transformation 
provides a simple method to understand dynamical localization
and recurrences in energy in kicked rotors.

In this mapping, the integer angular momentum quantum number of the rotor
corresponds to the lattice site in TBM. The free phase evolution
between kicks provides the pseudorandom diagonal (on-site) potential while 
the kicking potential $V(\theta)$ determines the range and strength of the
hopping. Thus, under certain conditions, boundedness and recurrences in energy
in the kicked rotor manifest themselves in quasienergy states which
are exponentially localized on a lattice labelled by the angular
momentum quantum number of the rotor.

>From a practical standpoint, this method of studying kicked rotors is
particularly useful when the TBM contains only nearest-neighbor couplings.
This has motivated many studies\cite{Fishman,Lloyd} of a special
class of kicked rotors where the potential is chosen to be
$V(\theta)= -2\hbar \arctan{(k cos(\theta))}$, resulting in a
TBM,
\begin{equation}
T_m u_m + \frac{k}{2} ( u_{m+1} + u_{m-1} ) = 0 \;.
\end{equation}
Note that we, unlike earlier treatments, explicitly retain the
presence of the quantization scale $\hbar$ in the definition of the
potential. This makes it clear why the classical limits of quantum
rotors corresponding to this choice of potential are trivial. The TBM, 
however, are perfectly well-defined and serve as useful illustrative
examples. 

This nearest-neighbor TBM has been studied~\cite{Fishman} for linear and quadratic rotors
which are described by $K(p)=\sigma p$, where $\sigma$ is an
irrational number, and
$K(p)= p^2/2$. In the diagonal term in the TBM, these translate into 
$K(m)=\sigma m$ and $K(m)=\hbar m^2/2$ respectively. Thus, $\hbar$
does not explicitly appear in the TBM analysis of the linear rotor. In 
the quadratic rotor we set, $\hbar = 8\pi\sigma$ in keeping with the
requirements of dynamical localization~\cite{Izrailev}.

The linear rotor, where the diagonal disorder is quasiperiodic,
was solved exactly. In particular, the
density of states was shown to be identical
to the average density of states for the Lloyd model of disorder~\cite{Lloyd}
for which the $T_m$ are independent random variables with
the Cauchy distribution
\begin{equation}
P(T_m) = \frac{1}{\pi} \frac{1}{1+T_m^2} \;,
\end{equation}
Furthermore, the numerically computed localization length was found to be
in good agreement with the 
analytic, 
ensemble averaged, inverse localization length $\bar{\gamma}$ of the Lloyd
model~\cite{Lloyd}, 
\begin{equation}
\cosh(\bar{\gamma}) =  \sqrt{k^{-2} + 1}\;.
\end{equation}
The localization
length $\gamma^{-1}$ depends upon $k$ and controls the transient
and recurrence time scales in the system.
For the quadratic rotor where the diagonal disorder is pseudorandom,
statistical studies of the sequences $T_m$ showed similarities with
the random potential with Lorentzian distribution. It was argued that
in spite of
some correlations, the quadratic model also exhibits the localization characteristics
of the Lloyd model~\cite{Fishman,Fishman2}. However, our work here
shows that {\it these residual correlations have a profound impact on the
fluctuations in the lineshape}. These, in turn, can lead to strong
deviations of $\gamma^{-1}$ from the ideal Lloyd model prediction. 

In this paper, we use a novel method to directly compute the fluctuations in the exponential
lineshape for linear and quadratic rotors. We compare and contrast
the linear, integrable, system 
with equally-spaced quasienergies, 
$\omega_j=j*\sigma$ mod(1) ( independent of $k$ ) and
the quadratic, nonintegrable case, where the quasienergies depend upon
$k$. We reiterate that the classical limit of the Lloyd model is 
not well-defined, and so integrability or not refers only to the
distribution of the energy levels. In other words, 
{\it the terminology merely distinguishes between the 
two cases we consider and does not imply a classical limit.}

Our motivation is to understand better how the two different $K(p)$ in the kicked rotor
are manifested in the localization properties of the equivalent TBMs.
We characterize the differences in these
two rotor systems by using a recent technique~\cite{KS} for studying the fluctuations in
the respective localized quasienergy states. We begin  
by factoring out the exponential envelope. Thus the projections of the
eigenstates of the rotor $u_m$ are written as 
\begin{equation}
u_m = e^{ -\gamma |m|} \eta_m \;,
\end{equation}
where $\eta_m$ are the fluctuations in the localized states at
the $m^{th}$ lattice site.
Thus, the fluctuations $\eta_m$ can be related to the fluctuations in the
asymptotic localization length $\bar{\gamma}$, (where
$u_m = e^{-\bar{\gamma}}$ ) as
\begin{equation}
\gamma - \bar{\gamma} = log|\eta_m|/m
\end{equation}
As explained below, we use an exact decimation scheme to compute the
scaling properties of the fluctuations
$\eta_m$ thereby establishing the fact that they are fractal.
Furthermore, the scale length
$\gamma^{-1}$ need not be presupposed but can be self-consistently determined.
It should be noted that this equation is valid for any TBM
including random systems and therefore the method described above can be
used to compute localization length for any nearest-neighbor TBM irrespective
of the nature of diagonal or off-diagonal disorder.

The fluctuations $\eta_m$ satisfy the following TBM
\begin{equation}
e^{-\gamma} \eta_{m+1}+e^{\gamma} \eta_{m-1}+ \lambda T_m \eta_m =0\; ,
\label{etatbm}
\end{equation}
where $\lambda = 2/k$.
We apply a recently developed decimation method
\cite{KS} to this TBM with
$\sigma$ in $T_m$ was taken to be
the inverse golden mean $(\sqrt 5 -1)/2$. 
In this approach, the incommensurability of the lattice was
exploited by decimating all
sites except those labelled by the Fibonacci numbers $F_n$.
This renormalization group approach was shown to an
extremely useful tool to demonstrate self-similarity and to obtain
universal characteristics of quasiperiodic systems.
Here, we apply this formalism to a linear rotor,
described by quasiperiodic TBM, as well as to the quadratic rotor
which is {\it not} quasiperiodic. We demonstrate that the Fibonacci decimation
scheme is a very efficient method to compute fluctuations
in the localization lengths irrespective of the nature of the aperiodicity
of the TBM. Note that for non-quasiperiodic problems, Fibonacci
decimation can be replaced by a more conventional one where every other
site is decimated.

After the $n^{th}$ decimation level, the nearest-neighbor TBM
connecting the fluctuations at two neighboring Fibonacci sites is
\begin{equation}
f_n(m) \eta(m+F_{n+1})=\eta(m+F_n) + e_n(m) \eta(m).
\end{equation}
The additive property of the Fibonacci numbers provide {\it exact} recursion
relations~\cite{KS}for the decimation functions $e_n$ and $f_n$:
\begin{eqnarray}
e_{n+1} (i)= - {A e_n (i) \over 1+Af_n (i)} \\
f_{n+1} (i)= {f_{n-1} (i+F_n) f_n(i+F_n)\over 1+Af_n(i)} \\
A = e_{n-1} (i+F_n) + f_{n-1} (i+F_n)e_n(i+F_n). \nonumber 
\end{eqnarray}
These can be iterated to machine precision
as they do not depend on any parameter which could limit the precision.
It turns out that for the localized phase where
$\gamma$ is always greater than zero, the
decimation function $f_n$ vanishes asymptotically
and hence the resulting renormalization flow simplifies to
\begin{equation}
e_{n+1}(i) = - e_{n-1}(i+F_n) e_n(i)\;.
\end{equation}
In view of this further simplification, the above equation can be iterated
up to $35$ iterations which corresponds to studying TBM of
size up to 14,930,352.

Any fractal character in the fluctuations can be inferred by non-trivial
asymptotic behavior of the functions $e_n$. In particular, the
convergence of the renormalization flow either to a 
non-trivial limit cycle ( which implies self-similarity ) or a strange set
is a clear indication of fractality.

It is interesting to note that the decimation
function $e_n$ also determines the localization length as
$\gamma - \bar{\gamma} = \log|e_n|/F_n$.
When the $\bar{\gamma}$ is not known, the above equation determines it
self-consistently to a very high precision.

In order to calculate the exponential lineshape for a given
quasienergy , we iterate the TBM (Eqn.(8)).
Starting from the site at $+N$, we iterate the 
equation backwards and simultaneously iterate forwards from site $-N$. We then match these
backward and forward iterates at $m=0$ by adjusting the phase factor $\omega$,
thereby determining the quasienergy.
Note that in this method, the localization center is
always at $m=0$.

An extremely accurate method to determine the quasienergies results
from rewriting the TBM as a quasiperiodically driven
map \cite{JKPhysicaD}. This is obtained by defining 
\begin{equation}
x_m = u_{m-1} / u_m \; ,
\end{equation}
which transforms the TBM to
\begin{equation}
x_{m+1} = {-1 \over x_m + \lambda \tan((\omega -K(m))/2)}\;.
\end{equation}
The localized phase of the TBM manifests itself as a strange nonchaotic
attactor of this map, reflected in the expression for
the Lyapunov exponent of the map $\mu = -2\gamma$~\cite{JKPhysicaD}.
The negative sign is crucial as it implies that an attractor with $\mu 
<0$ corresponds to diverging $u_m$ with increasing $m$. Similarly, an
attractor of the inverse map corresponds to diverging $u_m$ with
decreasing $m$. Thus, an exponentially localized function is obtained
by starting from the intersection points of these two attractors and
reversing the direction of iteration in each case.
The quasienergies of the kicked rotor are used as tunable phase
factors to ensure that {\it the intersection points of the attractor always 
occur} at $m=0$. This provides an extremely accurate method to compute the
quasienergies. The resulting $u_m$ is exponentially localized in both
directions with the localization center $m_0$
always at zero. Note that a consequence for the results we show is
that the lattice
site label is {\it always relative to the localization center}.

As stated earlier, the primary difference between 
the linear and quadratic rotors 
lies in the character of the fluctuations. In both cases,
the rotor wave function $u_m$, with exponentially decaying
envelope, exhibits fractal
fluctuations $\eta_m$ which decay as a power law, $\eta_m \approx m^p$ . The exponent
$p$ is related to the decimation function as
$p(n)=ln[abs(e_n)]/ln(F_n)$. Asymptotically ( in the
limit of large decimation level ), 
$e(n) \approx \eta(F_n)$, the exponent $p$ is a measure of 
the fluctuations in the exponent $\gamma$ as
$p(n)=(\gamma_n-\bar{\gamma})(F_n)/ln(F_n)$. Note that the measure $p(n)$
depends on the decimation level $n$. The behavior of $p$ with kicking
parameter $k$ distinguishes the two rotor cases we consider.

Figure~\ref{linfig} shows the
self-similar fluctuations
$\eta_m$, for $\omega=0$, in the case of the linear
rotor. It should be noted that the fluctuations are described by
a non-trivial period-$6$ limit cycle of the renormalization flow, that is 
$\eta_{F_n}=\eta_{F_{n+6}}$. Further, 
the same period-6 was found
for all values of $k$ thereby establishing the fact that the fractal
fluctuations are independent of $k$ for  $k > 0$. The multifractal nature
of these fluctuations was also confirmed by the $f(\alpha)$ curve.
For other values of quasienergies,
the fluctuations are not described
by a limit cycle. Instead,
the decimation functions $e_n$ (for all quasienergies) 
converge on an invariant set with fractal measure, independent of
$k$. For our purposes, it is more illustrative to focus on the impact of
these features on the localized lineshape. Given the relationships constructed earlier, 
this universality implies that {\it the convergence of the localization
length} $\gamma_n^{-1}$ 
to the asymptotic value $\bar{\gamma}^{-1}$ {\it is also independent}
of $k$.

As seen from the Fig.~\ref{flucfig}(a), the universality in the linear
rotor is reflected in $p(n)$ which is
independent of $k$ at different decimation levels. 
In contrast to the linear rotor, the quadratic rotor shows
fluctuations which depend strongly on $k$.
As seen from
Fig.~\ref{flucfig}(b), the power $p(n)$ exhibits many spikes as a
function of the parameter $k$. These spikes are the result of  
local "resonances" in the fluctuations. The nature of these resonances 
is seen from Fig.~\ref{flucfig}(c) where the magnitude of $\eta_m$
changes by several orders of magnitudes within a few sites. The location 
in $m$ and the magnitude of these `jumps' depend on both $k$ and the quasienergy
$\omega$. Two values of $k$ are shown in Fig~\ref{flucfig}(c) with the 
dark curve ($k=3.252$) corresponding to a larger spike (for both $n$)
in panel (b). Figure~\ref{flucfig}(d) shows that the resonances lead
to `shoulders' in the exponential lineshape, indicating local
variation in the exponential envelope. It is worth noting that the
quasienergies vary with $k$ and are computed for each $k$ using the
matching condition described earlier.

The location of the resonance determines the impact of these local deviations.
If the resonance occurs close to the localization center, then the
deviations in $\gamma$ from $\bar{\gamma}$ are clearly
significant. However, if the resonance
site is far from the localization center, then the exponential
envelope diminishes the importance of the deviations.
Figure~\ref{quasifig} shows the lineshapes associated with single
quasienergy states for both rotors.
As seen from the linear rotor
lineshape in Fig.~\ref{quasifig}(c), the {\it absence of
  resonances} means that the Lloyd
model estimate is a very good approximation to the actual calculated
localization length, {\it all the way to the localization center}.
This result was found to be true for all quasienergies and for
all values of the kicking parameter $k$.
However, for the quadratic rotor shown in Fig.~\ref{quasifig}(a) and (b), the 
Lloyd model estimate of $\gamma^{-1}$
appears to be correct only in the tails of the localized line shape for
most of the quasienergies.
Deviations from the asymptotic $\bar{\gamma}^{-1}$ depend strongly on
both the parameter $k$ and the quasienergy, as seen by contrasting
panel (a) and (b). For $\omega=0.17$, $\bar{\gamma}^{-1}$ is clearly
a better fit for the associated quasienergy state than for case of $\omega=0$.

\begin{figure}
\caption{Localized lineshapes (points) starting from a plane-wave initial
  condition for linear (panels (b) and (d)) and quadratic rotors ((a)
  and (c)). The first column considered $k=2.8$ while the second is
  for $k=5$. In all cases, the solid line indicates the Lloyd model result.}
\label{packevolv}
\end{figure}
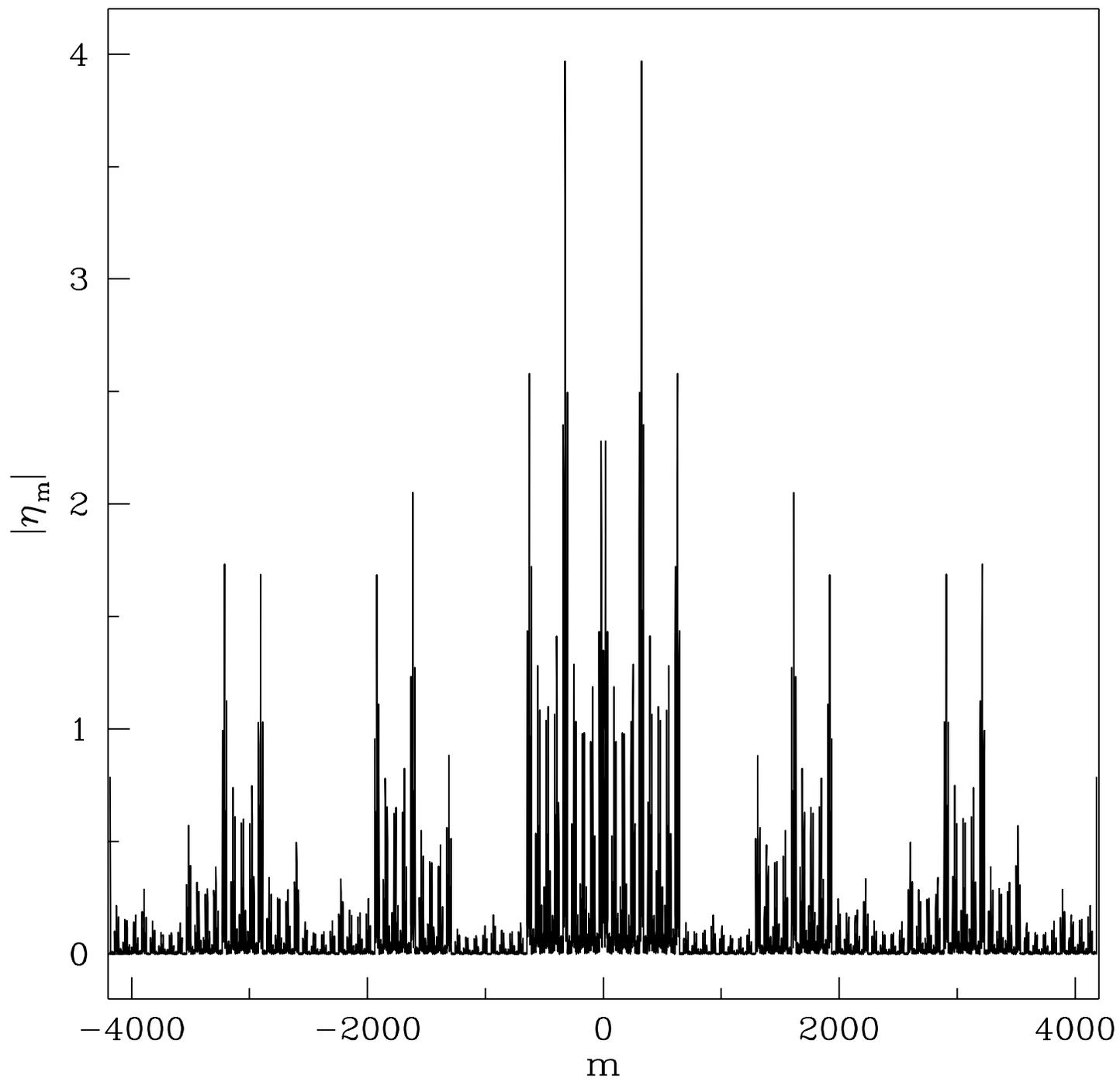
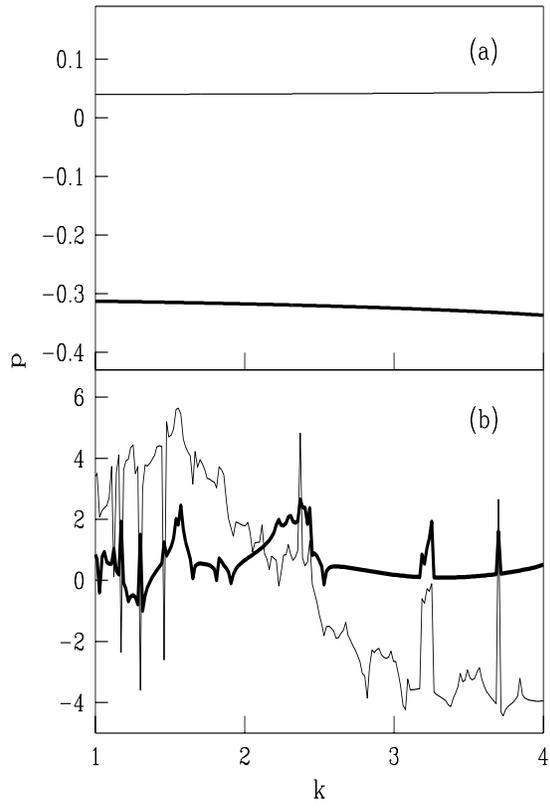
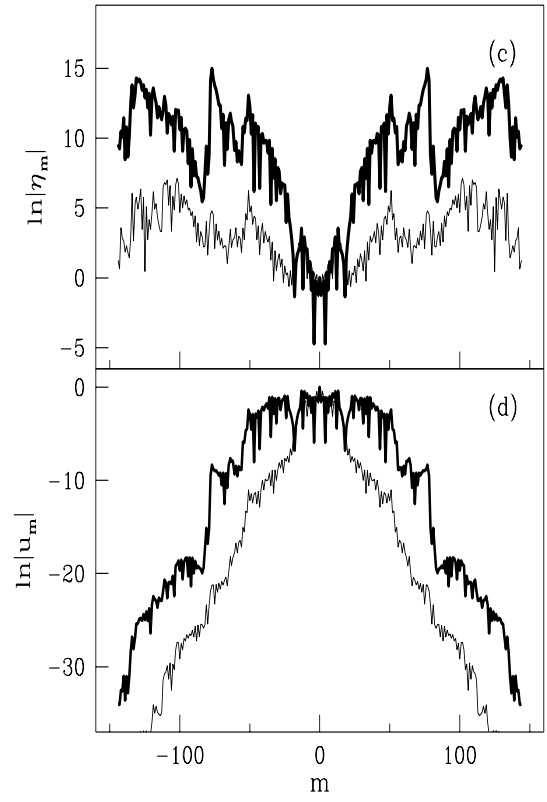
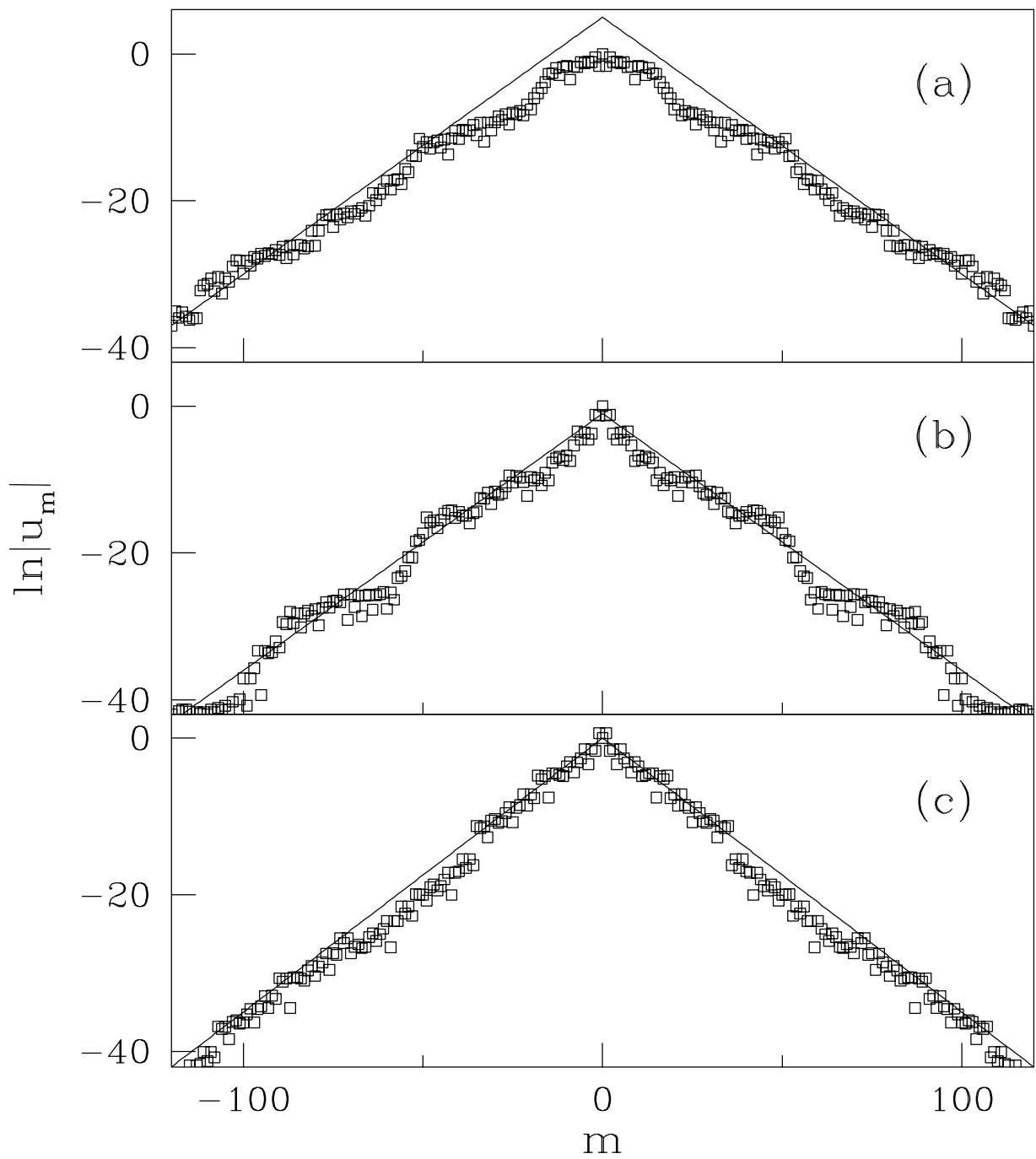
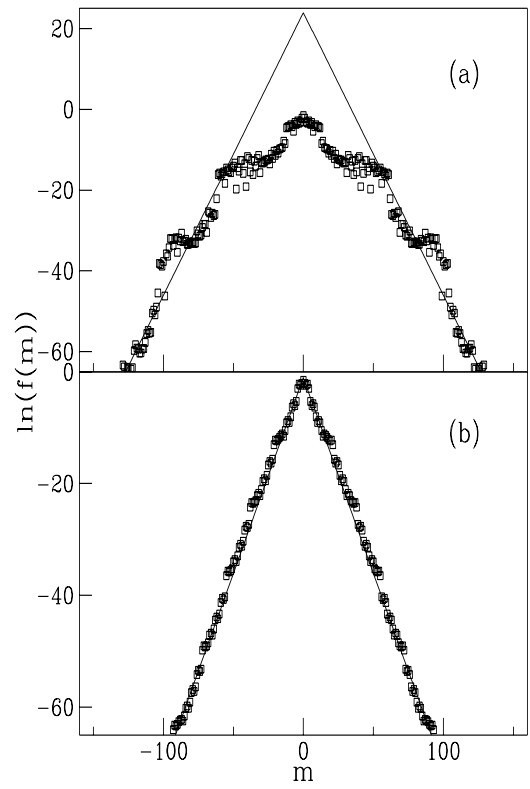
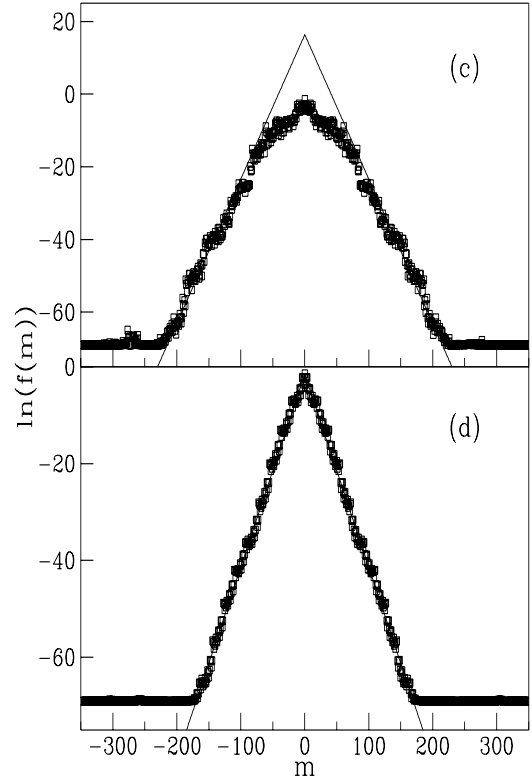

Therefore, localization in the quadratic rotor starting from an
initial wave packet composed of several quasienergy states should not be
well described by the Lloyd model. 
To illustrate this, we consider the evolution
of a plane wave (at $m=0$) under the repeated action of the
single-step evolution
operator $U$. The canonical method of Fast Fourier Transforms is used
to get the localized lineshape after a large number of kicks. The probability 
distribution $f(m)$ across lattice sites $m$ is then constructed.  As seen
from Fig~\ref{packevolv}, the linear rotor ((b) and (d)) coincides with the Lloyd
model prediction all the way to the localization center. This is in stark
contrast to the case of quadratic rotor in panels (a) and (c)  where the
deviations are strongest close to the localization center. It is also
evident that the magnitude of these deviations depend on $k$. Exponential 
fits near the centers of the lineshapes yield values of $\gamma^{-1}$
which are clearly different from the asymptotic estimates given by
$\bar{\gamma}^{-1}$.

The conditions under which the Lloyd model calculation~\cite{Lloyd} was made allow us
to speculate on the possible reasons for both the resonances and the
associated deviation in $\gamma$. The calculation of the estimate 
$\bar{\gamma}$ required
the $T_n$ to satisfy a specific distribution. However,
dynamical phase correlations would lead to a violation of this
requirement. This was verified both by constructing the return mapping
for the on-site potentials and by directly plotting a histogram of 
on-site terms and contrasting with the required
distribution. These were noted in earlier work as
well~\cite{Fishman,Fishman2} and are clearly different in the linear
and quadratic cases.

We believe that the local resonances and fluctuations seen in
the localization characteristics of the
quadratic rotor are generic to pseurorandom systems. 
We have verified that TBMs with bounded onsite potentials such as
$cos(2\pi \sigma m^{\nu})$ also exhibit
the characteristics similar to that of the quadratic rotor for $\nu > 2$. 
$\nu=1$ constitutes a special case as the model reduces to the
well-known quasiperiodic Harper equation~\cite{Harper}, where the
fractal localization character was also found to be independent of the
coupling~\cite{KS}. It should be noted that the Harper
equation has been recently solved using the
Bethe-Ansatz~\cite{BA} implying some sort of 'integrability' in the model.
In view of this, we speculate that for the linear rotor, the
$k$ independence results from the integrable nature of the problem. 

Ideally, in quadratic rotors, the possibility of correlations can be recognized from
studying the classical dynamics resulting from the kicked
rotor Hamiltonian. Specifically, in the context of quantum dynamics in
mixed phase spaces, quantum phase 
correlations can be associated with the presence of invariant
structures in the classical
phase space. This relationship is of great current interest in the new 
context of quantum manifestations of classical anomalous transport~\cite{BZ,Raizen}. 
Recent work suggests that there may be a relationship
between the large fluctuations
in the localization length and the anomalous diffusion in the classical 
phase space~\cite{BZ}. The lineshapes also exhibit shoulders similar
to the ones shown here. We propose to examine more closely the association of
quantum phase correlations with structures in the associated classical
phase space. However, as mentioned earlier, this is not
possible in the special class of rotor studied here as the
corresponding classical mapping is not well
defined. In keeping with this general motivation, we are presently extending our work to models
where this is not an issue.

\acknowledgements

The research of IIS is supported by a grant from National Science
Foundation DMR~093296. The work of BS was supported by the National Science
Foundation and a grant from the City University of New York
PSC-CUNY Research Award Program.

\begin{figure}
\caption{Self-similar fluctuations in a linear rotor for
$\omega=0$ . The fluctuations repeat themselves at every $6^{th}$ Fibonacci
site, that is $F(6),F(12),\cdots \mbox{etc.}$. This type of "translational invariance" in Fibonacci space
is described by a period-$6$ limit cycle of the renormalization
flow. Note that this behavior is independent of the kicking parameter
$k$.}
\label{linfig}
\end{figure}

\begin{figure}
\caption{ The variation in the power law exponent $p(n)$ as a function of the kicking
parameter $k$ is shown for (a) the linear and (b) quadratic rotors at
two different decimation levels. The darker and lighter curves respectively
correspond to $12^{th}$ and $15^{th}$ decimation level. Panel (c) shows the fluctuations
$\eta_m$ versus $m$ for the quadratic rotor at $k=3.252$ (darker curve) and at
$k=2.9$ (lighter curve). Note that the larger value of $k$ corresponds
to a peak (spike) in (b) while the other does not. The jump in
magnitude of the fluctuations is seen in the corresponding lineshapes
shown in (d), with larger fluctuations leading to `shoulders' seen in
the lineshape. These `shoulders' result in variations in localization
length from the corresponding asymptotic values. }
\label{flucfig}
\end{figure}

\begin{figure}
\caption{ Single localized quasienergy states (points) for
the quadratic rotor (a) and (b) and the linear rotor (c). The
parameter $k=2.8$. Note that
(a) and (b) correspond to quasienergies
0.00 and 0.17 respectively. The linear rotor case is computed for
$\omega=0$. The Lloyd model prediction for the localization length
$\bar{\gamma}^{-1}$ is plotted in each case (solid line) to guide the eye.}
\label{quasifig}
\end{figure}

\end{document}